\documentclass[twocolumn,showpacs,preprintnumbers,amsmath,amssymb, a4]{revtex4}

\usepackage{graphicx}
\usepackage{dcolumn}
\usepackage{amsmath}
\usepackage{multirow}
\usepackage{color}
\usepackage{bm}
\usepackage[nottoc]{tocbibind}
\usepackage{float}
\usepackage{enumitem}
\usepackage{amssymb}
\usepackage{mathtools}
\usepackage{tabularx}
\usepackage{subfig}

\usepackage{soul}
\usepackage{makecell}

\usepackage{hyperref}
\hypersetup{
  colorlinks   = true, 
  urlcolor     = blue, 
  linkcolor    = blue, 
  citecolor   = blue 
}

\DeclareMathOperator*{\argmin}{arg\min}
\newcommand{\ie}{\textit{i}.\textit{e}. \ }
\newcommand{\eg}{\textit{e}.\textit{g}. \ }

\begin{document}


\title{Committee machine that votes for similarity between materials}

\author{
Duong-Nguyen Nguyen$^{1}$, Tien-Lam Pham$^{1,2}$, Viet-Cuong Nguyen$^3$, Tuan-Dung Ho$^{1}$, Truyen Tran$^{4}$, Keisuke Takahashi$^{5}$, Hieu-Chi Dam$^{1,5,6, *}$   \\
}
\affiliation{
$^{1}$Japan Advanced Institute of Science and Technology, 1-1 Asahidai, Nomi, Ishikawa 923-1292, Japan\\
$^{2}$ESICMM, National Institute for Materials Science, 1-2-1 Sengen, Tsukuba, Ibaraki 305-0047, Japan\\
$^{3}$HPC Systems Inc., 3-9-15 Kaigan, Minato-ku, Tokyo 108-0022, Japan\\
$^{4}$Applied Artificial Intelligence Institute, Deakin University, Geelong, Australia. \\
$^{5}$Center for Materials Research by Information Integration, National Institute for Materials Science 1-2-1 Sengen, Tsukuba, Ibaraki 305-0047, Japan\\
$^{6}$JST, PRESTO, 4-1-8 Honcho, Kawaguchi, Saitama, 332-0012, Japan\\
$^*$dam@jaist.ac.jp
}

\date{\today}

\begin{abstract}

We developed a method for measuring the similarity between materials, focusing on specific physical properties. The obtained information can be utilized to understand the underlying mechanisms and to support the prediction of the physical properties of materials. The method consists of three steps: variable evaluation based on non-linear regression, regression-based clustering, and similarity measurement with a committee machine constructed from the clustering results. Three datasets of well-characterized crystalline materials represented by critical atomic predicting variables are used as test beds. Herein, we focus on the formation energy, lattice parameter, and Curie temperature of the examined materials. Based on the information obtained on the similarities between the materials, a hierarchical clustering technique is applied to learn the cluster structures of the materials that facilitate interpreting  the mechanism, and  an improvement of regression models is introduced for predicting the physical properties of the materials. Our experiments show that rational and meaningful group structures can be obtained and that the prediction accuracy of the materials’ physical properties can be significantly increased, confirming the rationality of the proposed similarity measure.

\end{abstract}

\maketitle

\section{Introduction \label{introduction}}
Computational materials science encompasses a range of methods that are used to model materials and simulate their responses on different length and time scales \cite{BGSumpter2015}. The majority of problems addressed by computational materials science are related to methods that focus on two central tasks. The first aims to predict the physical properties of materials, and the second aims to describe and interpret the underlying mechanism \cite{Liu2017, Lu2017, Zachary2017}. In the first task of predicting physical properties, computer-based quantum mechanics techniques \cite{Jain2016, KohnSham65, JonesGunnarsson89, Jones2015} in the form of well-established first-principles calculations are generally performed with high accuracy and are applicable to any material, but with high computational cost. Recently, the increase in advanced machine learning techniques \cite{ML, Tib2009, TuLe2012} and volume of computational material databases \cite{Jain2013,  OQMD2013} has provided new opportunities for researchers to automatically construct prediction models (from a huge amount of precomputed data) that predict specific physical properties with the same level of high accuracy, while dramatically reducing the computational costs \cite{Behler07, Rupp12, Pilania2013, Fernandez2014, Smith17}.

By contrast, the second task, \ie describing and interpreting the mechanisms underlying the physical properties of materials, relies mostly on the experience, insight, and even luck of the experts involved. In fact, comprehension of multivariate data with nonlinear correlations is typically extremely challenging, even for experts. Thus, the utilization of data mining and machine learning techniques to discover hidden structures and latent semantics in multidimensional data \cite{Lum2013, LSA1998, Blei2012} of materials is promising, but only limited works have been reported so far \cite{Kusne15, Srikant2015, Scheffler117}. 

To apply well-established machine-learning methods to solve problems in materials science, the primitive representation of materials must usually be converted into vectors, such that the comparison and calculations using the new representation reflect the nature of the materials and the underlying mechanisms of the chemical and physical phenomena. However, real-world applications, especially for solving the second task, often focus on physical properties of which the mechanism is not fully understood \cite{Krishna2015, Scheffler2015}. In these cases, it is almost impossible to appropriately represent the materials as vectors of features so that comparisons using well-established mathematical calculations can reflect the similarity/dissimilarity between them. Therefore, a true data-driven approach for solving materials science problems still requires much further fundamental development.

In this study, we focus on establishing a data-driven protocol for solving the second task of computational materials science. Focusing on a specific physical property, we aim to develop a method for measuring the similarity between materials from the viewpoint of the underlying mechanisms that work in these materials. The method for measuring this similarity consists of three steps: (1) variable evaluation based on non-linear regression, (2) regression-based clustering, and (3) similarity measurement with a committee machine \cite{Tresp2000, Opitz1999} constructed based on the clustering results. The variable evaluation \cite{Liu05, Blum97} aims to identify and remove unneeded, irrelevant, and redundant variables from the data \cite{Rakkrit09, Almuallim91, Biesiada07}. We carried out this analysis in an exhaustive manner by testing all combinations of predicting variables to find the variables with the potential to yield good prediction accuracy $(PA)$ for the target variable. The regression-based clustering method is developed from the well-known K-means clustering method \cite{LLOYD82, macqueen1967, Kanungo02} with major modifications for breaking down a large dataset into a set of separated smaller datasets, in each of which the target variables can be predicted by a different linear model. Regression-based clustering models are then constructed for all the selected potential combinations of predicting variables, so as to construct a committee machine that votes for the similarity between the materials.

Three datasets of well-characterized crystalline materials represented by appropriate predicting variables, together with their physical properties as determined through first-principles calculations or measured experimentally, are used as test beds. Our experiments show that the proposed similarity measure can derive rational and meaningful material groupings and  can significantly improve the $PA$ of the physical properties of the examined materials.

\section{Methods}\label{AlgorithmOuline}
\begin{figure*}[t]
\centering
\includegraphics[scale=0.50]{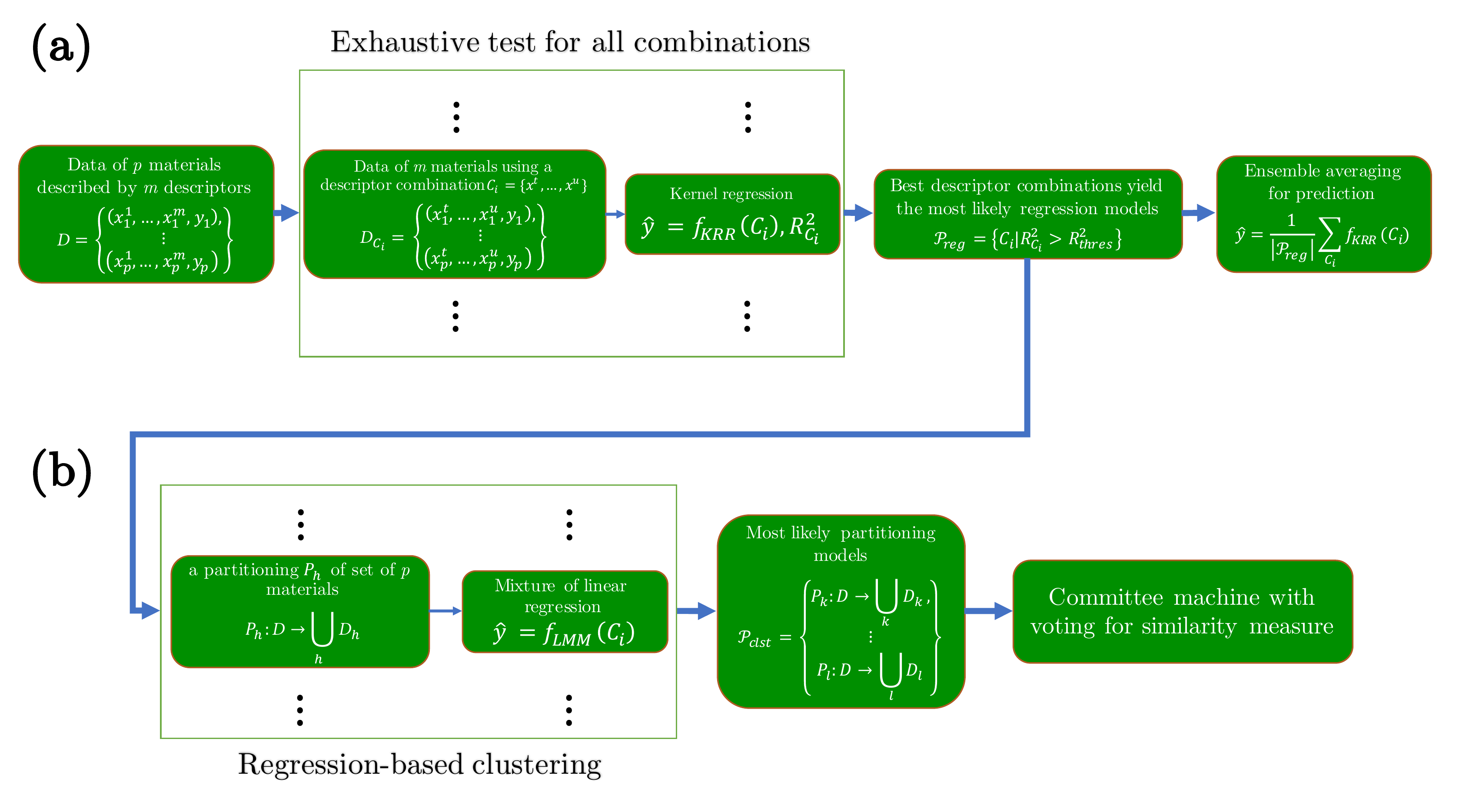}
\caption{The data flow in our proposed method to measure similarity between materials, focusing on specific target physical properties and using MapReduce representation language. The process consists of two sub-processes: (a) an exhaustive test for all predicting variable combinations from which we can select the best combinations yielding the most likely regression models, and (b) a utilization of the regression-based clustering technique to search for partition models that can break down the dataset into a set of separated smaller datasets, so that each target variable can be predicted by a different linear model. We can obtain a prediction model with higher $PA$ by taking an ensemble average of the yielding models in (a). We use the obtained partitioning models in (b) to construct a committee machine that votes for the similarity between materials.}
\label{FlowChart}
\end{figure*}

We consider a dataset $\mathcal{D}$ of $p$ materials. Assume that a material with index $i$  is described by an $m$-dimensional predicting variable vector ${\bm{x}}_{i} = {\left( {x_i^1}, {x_i^2}, \dots , {x_i^m} \right)} \in {\mathbb{R}}^m$. The dataset $\mathcal{D}$ is then represented using a $(p \times m)$  matrix. The target physical property values of the materials are stored as a $p$-dimensional target vector $ \bm{y} = \left( y_1, y_2 \dots y_p \right) \in \mathbb{R}^p$. The entire data analysis flow is shown in Figure \ref{FlowChart}. 

\subsection{Kernel regression-based variable evaluation}\label{NLFS}
To develop a better understanding of the processes that generated the data, we first utilize an exhaustive search to evaluate all variable combinations \cite{Kohavi1997, Liu05, Blum97} to identify and remove unneeded, irrelevant, and redundant variables \cite{Rakkrit09, Almuallim91, Biesiada07}. We begin by learning non-linear functions for predicting the values of a specific physical property (target quantity) of the materials. We apply the Gaussian kernel ridge regression (GKR) technique  \cite{ML}, which has recently been applied successfully to several challenges in materials science  \cite{Rupp_tutorial, Botu, Pilania}. For GKR, the predicted property $y=f(\bm{x})$ at a point $\bm{x}$ is expressed as the weighted sum of Gaussians:

\begin{equation}
f(\bm{x}) = \sum_{i=1}^{p} c_{i} \exp \left( \frac{ {-{||{\bm{x}}_{i} - \bm{x}||}_{2}^{2}} }{2 \sigma^{2}} \right)
\end{equation}

 where $p$ is the number of training data points, $\sigma^2$ is a parameter corresponding to the variance of the Gaussian kernel function, and ${{||{\bm{x}}_i - \bm{x}||}_2^2} = \sum_{\alpha = 0}^{m} {\left( x_i^{\alpha} - x^{\alpha}  \right)}^2 $ is the squared $L^2$ norm of the difference between the two $m$-dimensional vectors $\bm{x}_i$ and $\bm{x}$. The coefficients $c_i$ are determined by minimizing

\begin{equation}
\sum_{i=1}^{p} {\left[ f(\bm{x_{i}}) - y_{i} \right] }^{2} +\lambda \sum_{i=1}^{p} {||c_{i}||}_{2}^{2} 
\end{equation}
where $y_{i}$ is the observed data value for data instance $i$. The regularization parameters $\lambda$ and $\sigma$ are selected with the help of cross-validation, \ie by excluding some of the materials as a validation set during the training process and measuring the coefficient of determination $R^2$, which is defined \cite{Kvalseth85} as follows:

\begin{equation}
R^2 = 1 - \frac {\sum_{j=1}^{p_{vld}} {\left[ f(\bm{x}_j) - y_j \right]}^2} {\sum_{j=1}^{p_{vld}} {\left[ \bar{y} - y_{j} \right]}^2}
\end{equation}
Here, $p_{vld}$ is the number of data points and $\bar{y}$ is the average of the validation set used to compare the values predicted for the excluded materials with the known observed values. In this study, we use $R^2$ as a measure of $PA$. 

To accurately estimate $PA$, we cross-validate the GKR \cite{Stone74, Picard84,  Kohavi95astudy} using the collected data repeatedly. To obtain a set of proper variable combinations that can accurately predict the target variable, we train the GKR models for all possible combinations of numerical predicting variables. It should be noted that, since we do not know yet the effect of each predicting variable on the target quantity, all the numerical predicting variables are normalized in the same manner in this analysis. With each combination, we search for the regularization parameters to maximize $PA$ of the corresponding GKR model. Note that  each of the selected combinations contributes a perspective on the correlation between the target and the predicting variables. Thus, an ensemble averaging \cite{Tresp2000, Dietterich2000, Zhang2012} technique can be applied to combine all the pre-screened regression models to improve the $PA$. Further, the similarity between materials regarding the mechanism of the chemical and physical phenomena associated with the target quantity can be investigated more comprehensively if we consider all the perspectives. Consequently, we need to construct regression-based clustering models for each obtained potential combination to build the committee machine.

\subsection{Regression-based clustering}\label{RBC}
In practice, a single linear model is often severely limited for modeling real data, because the data set can be non-linear or the data itself can be heterogeneous and contain multiple subsets, each of which fits best with a different linear model. However, in traditional data analysis, linear models are often preferred because of their highly advantageous interpretability. Within a linear model, one can intuitively understand how the predicting variables contribute to the target variable. Therefore, several efforts have been devoted to developing subspace segmentation techniques to decompose a high-dimensional dataset into a set of separate small datasets, each of which can be approximated well by different linear subspaces by employing principal component analysis \cite{Fukunaga1971, Vidal2005, Einbeck2008}. 

In this study, our primary interest is the local linearity between the predicting variables and the target variable, which may reflect the nature of the underlying physics around the point of observation. Therefore, we employ a simple strategy, in which the subspace segmentation is an integration of a conventional clustering method and linear regression analysis. It should be noted that the subspaces may have fewer dimensions than the whole space. Hence, we apply the sparse linear regression analysis using $L1$ regularization \cite{Tibshirani1996} instead of the original one.

Our proposed regression-based clustering method is based on the well-known K-means clustering method with two modifications. (1) The sparse linear regression model derived from data associated with materials in a particular cluster (group) is considered to be its common characteristic (center). The dissimilarities in the characteristics of each material in a group relative to the shared (common) nature of that group (the distance to the center) are measured according to its deviation from the corresponding linear regression model. (2) The sum of the differences of all materials in a group from the corresponding linear regression model of another group is used to measure the dissimilarity in the characteristics of that group with regard to the other group. The sum of the dissimilarities between one group and another and that determined in the reverse direction are used to assess the divergence between the two groups. 

After performing the variable evaluation, we assume we have selected combinations of predicting variables that yield non-linear regression models of high $PA$. With one of the selected combinations,  $m'$ numerical variables are selected from the original $m$ numerical variables. A material in the dataset is then described by an $m'$-dimensional predicting variable vector 
$\bm{x}'_{i} = {( {x_i^1}, {x_i^2}, \dots , {x_i^{m'}} )} \in {\mathbb{R}}^{m'}$, and the data are represented using a  $(p \times m')$  matrix.

Given the set  $\mathcal{D}$ of $p$ data points represented by $m'$-dimensional numerical vectors, a natural number $k \leq p$ represents the number of clusters for a given experiment. We assume that there are $k$ linear regression models and that each data point in $\mathcal{D}$ follows one of them. The aim is to determine those $k$ linear regression models, accordingly, to divide $\mathcal{D}$ into $k$ non-empty disjoint clusters. Our algorithm searches for a partition of $\mathcal{D}$ into $k$ non-empty disjoint clusters $({\mathcal{D}}_{1}, { \mathcal{D}}_{2},  \dots , {\mathcal{D}}_{k})$ that minimize the overall sum of the residuals between the observed and predicted values (using the corresponding models) of the target variable. The problem can be formulated in terms of an optimization problem as follows.

For a given experiment with cluster number $k$, minimize
\begin{equation}\label{equation1}
P(W, M) = \sum_{i=1}^{k} \sum_{j=1}^{p} w_{ij} \parallel y_{j} - y_{j}^{M_{i}}\parallel
\end{equation}
subject to
\begin{eqnarray}\label{equation2}
& \forall j: \sum_{i=1}^{k} w_{ij} = 1,  w_{ij} \in \{0, 1\}  \\
&1 \leq k \leq p, 1 \leq i \leq k, 1 \leq j \leq p
\end{eqnarray}
where $y_{j}$ and  $y_{j}^{M_{i}}$ are the observed value and the value predicted by model $M_{i}$ (of $k$ models) for the target property of the material with index $j$; $W = {\left[ w_{ij} \right]}_{p \times k}$ is a partition matrix ($ w_{ij}$ takes a value of 1 if object $x_{j}$ belongs to cluster $\mathcal{D}_{i}$ and $0$ otherwise), and $M = \left( M_{1}, M_{2}, \dots , M_{k} \right)$ is the set of regression models corresponding to clusters $({\mathcal{D}}_{1}, { \mathcal{D}}_{2},  \dots , {\mathcal{D}}_{k})$.

$P$ can be optimized by iteratively solving two smaller problems:

\begin{itemize}
\item fix $M = \hat{M}$ and solve the reduced problem $P(W, M)$ to find $\hat{W}$ (re-assign data points to the cluster of the closest center);
\item fix $W = \hat{W}$ and solve the reduced problem $P(W, M)$ to find $\hat{M}$ (reconstruct the linear model for each cluster).
\end{itemize}

Our regression-based clustering algorithm comprises three steps and iterates until $P(W, M)$ converges to some local minimum values:

\begin{enumerate}
\item The dataset is appropriately partitioned into $k$ subsets, $1\leq k \leq p$. Multiple linear regression analyses are independently performed with the $L1$ regularization method \cite{Tibshirani1996} on each subset to learn the set of potential candidates for the sparse linear regression models $M^{(0)} = \left\{  M_{1}^{(0)}, M_{2}^{(0)}, \dots , M_{k}^{(0)} \right\}$. This represents the initial step $t = 0$;

\item $M^{(t)}$ is retained and problem $P(W, M^{(t)})$ is solved to obtain $W^{(t)}$, by assigning data points in $\mathcal{D}$ to clusters based upon models  ${M_{1}^{(t)}, M_{2}^{(t)}, \dots , M_{k}^{(t)}}$;

\item $W^{(t)}$ is fixed and $M^{(t)}$ is generated such that $P(W, M^{(t+1)})$ is minimized. That is, new regression models are learned according to the current partition in step 2. If the convergence condition or a given termination condition is fulfilled, the result is output, and the iterations are stopped. Otherwise, $t$ is set to $t+1$ and the algorithm returns to step 2.
\end{enumerate}

The group number $k$ is chosen considering two criteria: high linearity between the predicting and target variables for all members of the group, and no model representing two different groups. The first criterion has higher priority and can be  quantitatively evaluated by using the Pearson correlation scores between the predicted and observed values for the target variable of the data instances in each group, by applying the corresponding linear model. The second criterion is implemented to avoid the case in which one group with high linearity is further divided into two subgroups that can be represented by the same linear model. The determination of $k$, therefore, can be formulated in terms of an optimization problem, as follows:

\begin{equation}
k = \argmin_{k \leq p} \left[\log\frac{1-\min_{1 \leq i \leq k} R^2_{i,i} }{\min_{1 \leq i \leq k} R^2_{i,i}}+ \max_{1 \leq i\neq j\leq k } R^2_{i,j} \right]
\label{clustering_criteria}
\end{equation}
where $R^2_{i,i}$ and $R^2_{i,j}$ are the Pearson correlation scores between the predicted and observed values for the target variable when we apply the linear model $M_i$ to data instances in clusters $i$ and $j$, respectively. 

The first term in this optimization function monotonically decreases with respect to the range of $\min_{1 \leq i \leq k} R^2_{i,i}$ varying from 0 to 1. When $\min_{1 \leq i \leq k} R^2_{i,i}$ approaches 1 (the entire cluster exhibits almost perfect linearity between the target and predicting variables), the optimization function drops on a $\log$ scale to emphasize the expected region. In contrast, the optimization function exponentially increases when $\min_{1 \leq i \leq k} R^2_{i,i}$ approaches 0 (one of the clusters shows no linearity between the target and predicting variables). The second term in this optimization function is introduced to avoid overestimation of $k$, in which a group with high linearity further divides into two subgroups that can be represented by the same linear model. It should be noted that the criterion for determining $k$ is also the criterion for evaluating a regression-based clustering model. Further, cluster labels can be assigned for a material  without knowing the value of the target physical property, by using the estimated value obtained from a prediction model, \eg a non-linear regression model.

\subsection{Similarity measure with committee machine}\label{SimilarityMap}
A clustering model, obtained through regression-based clustering for a particular combination of predicting variables, represents a specific partitioning of the dataset into groups in which the linear correlations between the predicting and target variables can be observed. The materials belonging to the same group potentially have the same actuating mechanisms for the target physical property. However, materials that actually have the same actuating mechanisms for a specific physical property should be observed similarly in many circumstances. Therefore, the similarity between materials, focusing on a specific physical property, should be measured in a multilateral manner. For this purpose, for each pre-screening of the sets of predicting variables that yield non-linear regression models of high $PA$ (section \ref{NLFS}), we construct a regression-based clustering model. A committee machine that votes for the similarity between materials is then constructed from all obtained clustering models. The similarity between two materials can be measured naively using the committee algorithm \cite{Seung1992, Settles2010}, by counting the number of clustering models that partition these two materials into the same cluster. The affinity matrix $A$ of all pairs of materials in the dataset is then constructed as follows:

\begin{equation}\label{AffinityMatrix}
A_{a,b} = \frac{1}{|S_h|} \sum_{\forall{S} \in S_h} \sum_{i=1}^{k_{S}} w^{S}_{ia} w^{S}_{ib} 
\end{equation}
 where $S_h$ is the set of all pre-screened combinations of  predicting variables that yield non-linear regression models of high $PA$ and $k_s$ is the cluster number. Further, $W^{S} = {\left[ w^{S}_{ij} \right]}_{p \times k_S}$ is the partition matrix of the clustering models obtained through the regression-based clustering analysis using the combination of predicting variables $S$ ($ w^{S}_{ia}$ takes a value of 1 if material $a$ belongs to cluster $i$ and $0$ otherwise). By using this affinity matrix, one can easily implement a hierarchical clustering technique \cite{HAC} to obtain a hierarchical structure of groups of materials that have similar correlations between the predicting and target variables.

\section{Results and discussion}\label{ResultAndDiscussion}
We applied the methods described above to a sequential analysis for automatic extraction of physicochemical information relating to considered materials from three available datasets. For each dataset, a brute force examination of all combinations of numerical predicting variables was conducted using a non-linear regression technique, to identify combinations of predicting variables that yielded regression models of high $PA$ for the later analysis process. For each of the pre-screened combinations, physically meaningful patterns in the form of material groups, as well as the linear relationships between the selected predicting and target variables, could be detected automatically for the materials in each group utilizing the regression-based clustering technique. The committee machine was then constructed from the obtained clustering models. Subsequently, a hierarchical structure of material groups similar to each other could be extracted using the hierarchical clustering technique. We evaluated the obtained results from both qualitative and quantitative perspectives. The qualitative evaluations were based on the rationality and interpretability of the obtained hierarchy with reference to the domain knowledge; the quantitative evaluations were performed based on the $PA$ of the predictive models constructed with reference to the obtained similarity between materials.

\subsection*{Experiment 1: Mining quantum calculated formation energy data of $\textit{Fm}{\bar{3}}\textit{m}$ AB materials}\label{secABcompound}

In this experiment, we collected computational data for 239 binary AB materials from the Materials Project database \cite{Jain2013}. The A atoms were virtually all metallic forms: alkali, alkaline earth, transition, and post-transition metals, as well as lanthanides. The B elements, by contrast, were mostly all metalloids and non-metallic atoms. We set the computed formation energy $E_{form}$ of each AB material as the physical property of interest. To simplify the demonstration of our method, we limited the collected compounds to those possessing the same cubic structure as the $\textit{Fm}{\bar{3}}\textit{m}$ symmetry group (\ie the NaCl structure).

To represent each material, we used a set of 17 predicting variables divided into three categories, as summarized in Table \ref{ABVarTable}. The first and second categories pertained to the predicting variables of the atomic properties of the element A and element B constituents; these included eight numerical predicting variables: (1) atomic number ($Z_A$, $Z_B$); (2) atomic radius ($r_A$, $r_B$); (3) average ionic radius ($r_{ionA}$, $r_{ionB}$), (4) ionization potential ($IP_A$, $IP_B$); (5) electronegativity ($\chi_A$, $\chi_B$); (6) number of electrons in outer shell ($n_{eA}$, $n_{eB}$); (7) boiling temperature  ($T_{bA}$, $T_{bB}$); and (8) melting temperature  ($T_{mA}$, $T_{mB}$) of the corresponding single substances. The boiling and melting temperatures were as measured under standard conditions ($0^{\circ}$C, $10^5$ Pa). 

Information related to crystal structure is very valuable for understanding the physical properties of materials. Therefore, we designed the third category with structural predicting variables whose values were calculated from the crystal structures of the materials. In this experiment, owing to the similarities in the crystal structures of the collected materials, we utilized only the unit cell volume ($V_{cell}$) as the structural predicting variable. The computed $E_{form}$ of each material was set as the target variable. 

\begin{figure*}[t]
\centering
\includegraphics[scale=0.28]{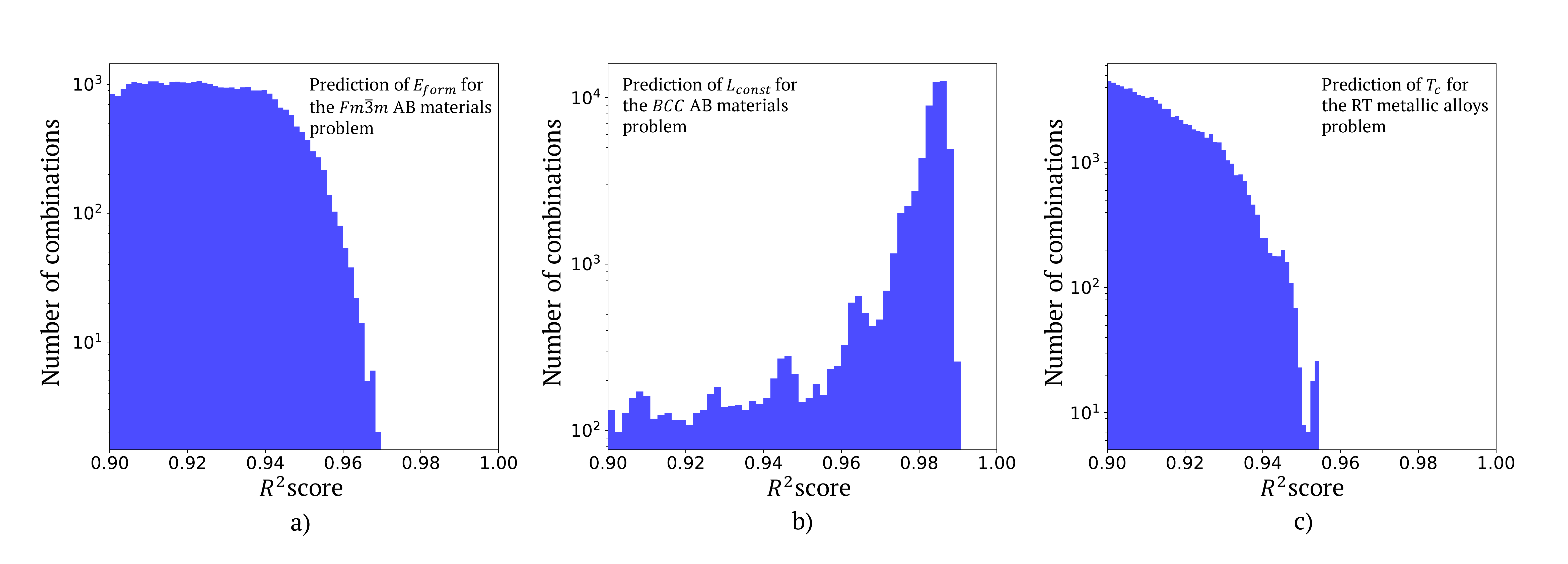} 
  \caption{Number of predicting variable combinations that yield corresponding prediction models  with $R^2$  larger than 0.90 for different problems: a) prediction of $E_{form}$ for the $\textit{Fm}{\bar{3}}\textit{m}$ AB materials, b) prediction of $L_{const}$ for the $BCC$ AB materials, and c) prediction of $T_{c}$ magnetic phase transition temperature for the rare-earth--transition metal alloys.}
  \label{NLFS_scores}
\end{figure*}

\begin{table}
\centering
\caption{Designed predicting variables describe intrinsic properties of constituent elements and structure-properties of materials in $E_{form}$ prediction problem. The A and B elements comprise the AB materials with binary cubic structure identical to that of the $\textit{Fm}{\bar{3}}\textit{m}$ symmetry group. }
\label{ABVarTable}
\begin{tabular}{p{2.6cm}p{5.5cm}}
\hline\hline
Category &  Predicting variables
\\
\hline
Atomic properties of A element& $Z_A$, $r_{ionA}$, $r_A$, $IP_A$, $\chi_A$,  $n_{eA}$, $T_{bA}$, $T_{mA}$\\
\hline
Atomic properties of B element& $Z_B$, $r_{ionB}$, $r_B$, $IP_B$, $\chi_B$,  $n_{eB}$, $T_{bB}$, $T_{mB}$\\
\hline
Structural information &$V_{cell}$     
 \\
\hline\hline       
\end{tabular}
\end{table}

A kernel regression-based variable evaluation was performed for these data with 3-times 10-fold cross-validations. We first examined how $E_{form}$ can be predicted from the designed predicting variables for all collected materials. We performed a screening for all possible ($2^{17}$ - 1 = 131,071) variable combinations. Hence, we found a total of 34,468 variable combinations deriving GKR models with $R^2$ scores exceeding 0.90 (Fig$.$\ref{NLFS_scores}). Among them, there were 139 variable combinations deriving GKR models with $R^2$ scores exceeding $0.96$. These predicting variable combinations were then considered as candidates for the next step of the analysis. The highest prediction accuracy $PA$ in this experiment is 0.967 (mean of absolute error, abbreviated as MAE: 0.122 eV), obtained using the combination $\left\{V_{cell}, {\chi}_{A}, n_{eA}, n_{eB}, {IP}_{A}, T_{bA}, T_{mA}, r_{B}\right\}$. Moreover, we could obtain superior $PA$ with an $R^{2}$ score of 0.972 (MAE: 0.117 eV) by taking ensemble averages \cite{Tresp2000, Dietterich2000, Zhang2012} of GKR models, which were constructed using the 139 selected variable combinations.

\begin{figure*}[t]
\centering
  \includegraphics[scale=0.31]{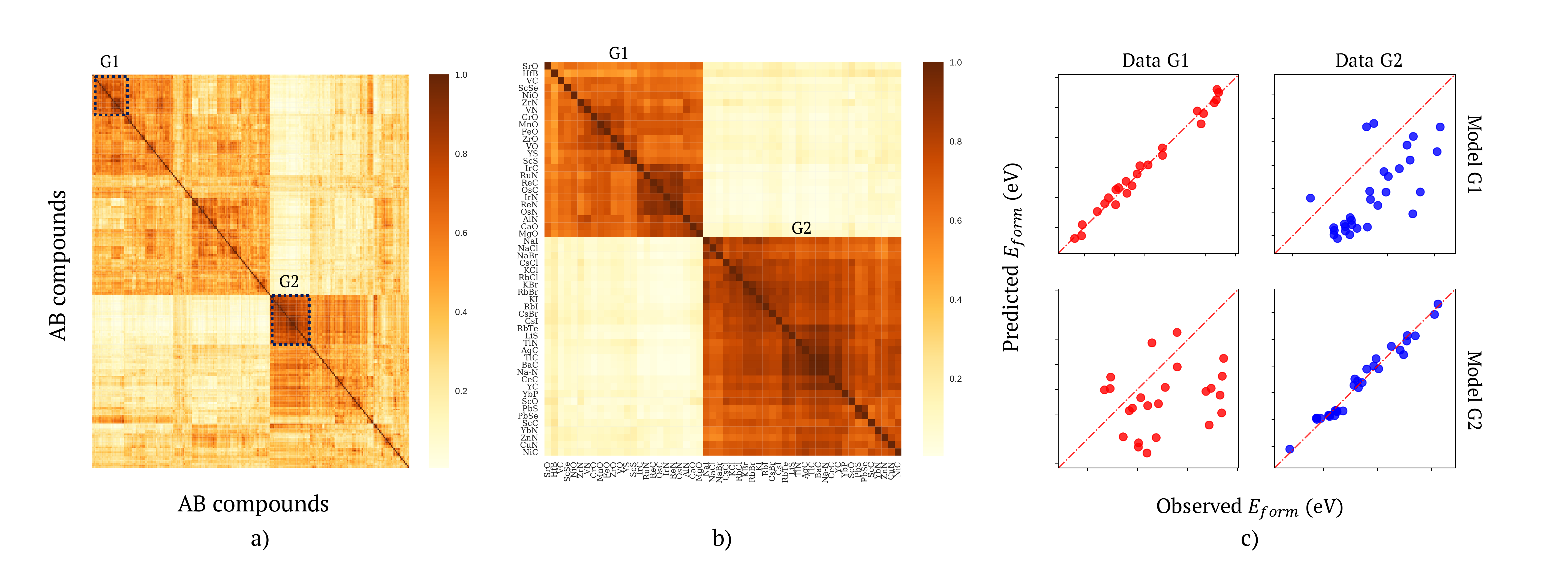}
  \caption{a) Affinity matrix between the $\textit{Fm}{\bar{3}}\textit{m}$ AB materials yielded by regression-based committee voting machine. b) Enlarged view of highly similar elements in G1 and G2 regions in affinity matrix. c) Confusion matrixes measuring linear similarities among materials in G1 and G2, as well as dissimilarities between models generated for materials in different groups.}
  \label{ABmap}
\end{figure*}

We performed regression-based clustering analyses for all 139 selected variable combinations with 1000 initial randomized states. By using evaluation criteria similar to those for determining the number of clusters (formula \ref{clustering_criteria}), the 200 best clustering results among these trials were selected to construct a committee machine that voted for the similarity between materials. The obtained affinity matrix for all the $\textit{Fm}{\bar{3}}\textit{m}$  AB materials is shown in Fig$.$\ref{ABmap}a. The similarity between each material pair varies from 0 to 1. A cell of the affinity matrix takes a 0 value when the corresponding two materials are never included in the same cluster by a regression-based clustering model. In contrast, a cell of the affinity matrix takes a value of 1 when the corresponding two materials always appear in the same cluster according to every regression-based clustering model. By using this similarity, we could roughly divide all the materials into two groups, as represented by the upper left and bottom right of Fig$.$\ref{ABmap}a. 

\begin{figure*}
\centering
\subfloat[][] {{\includegraphics[scale=0.27]{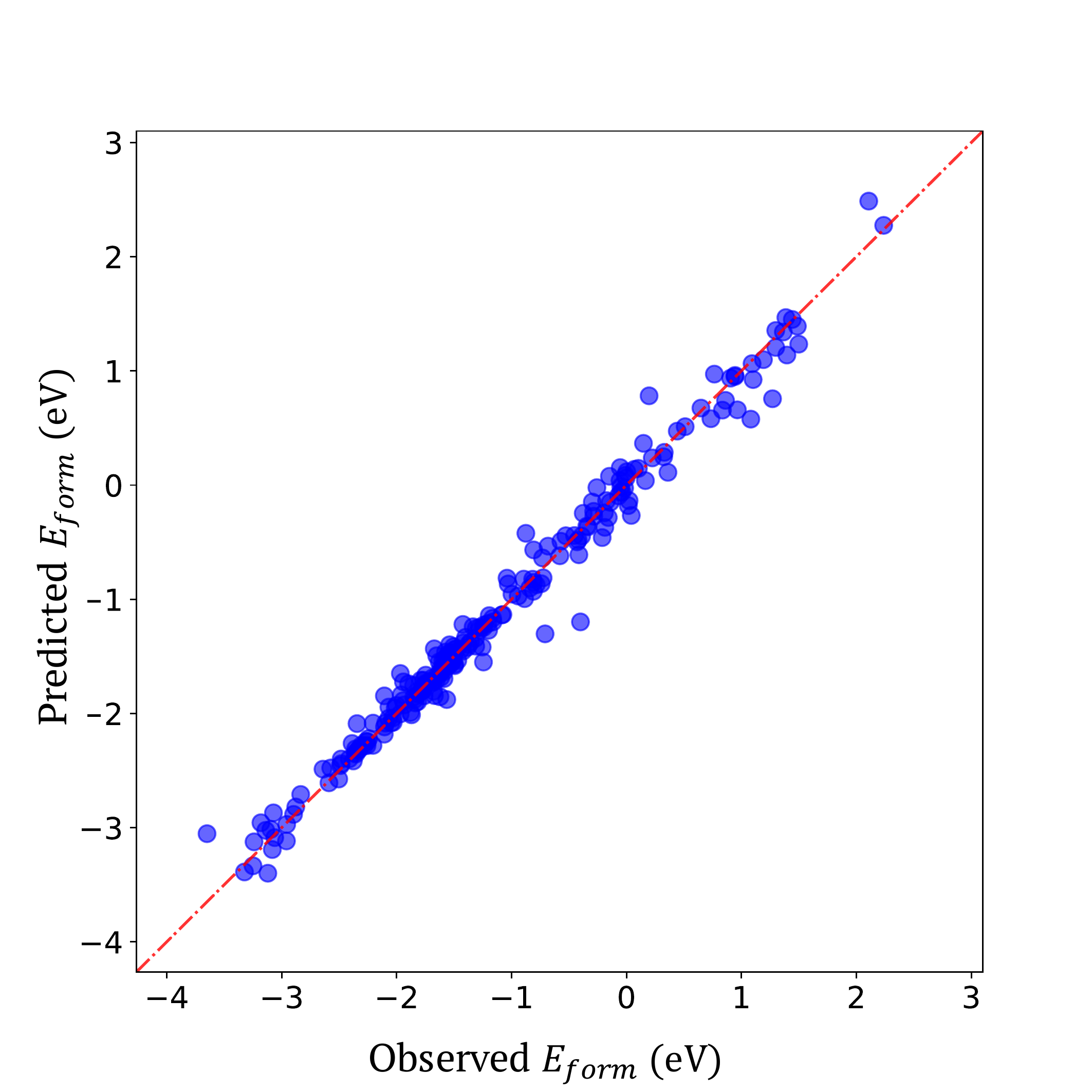} } \label{NlfsBestAB}}
 \subfloat[][] {{\includegraphics[scale=0.27]{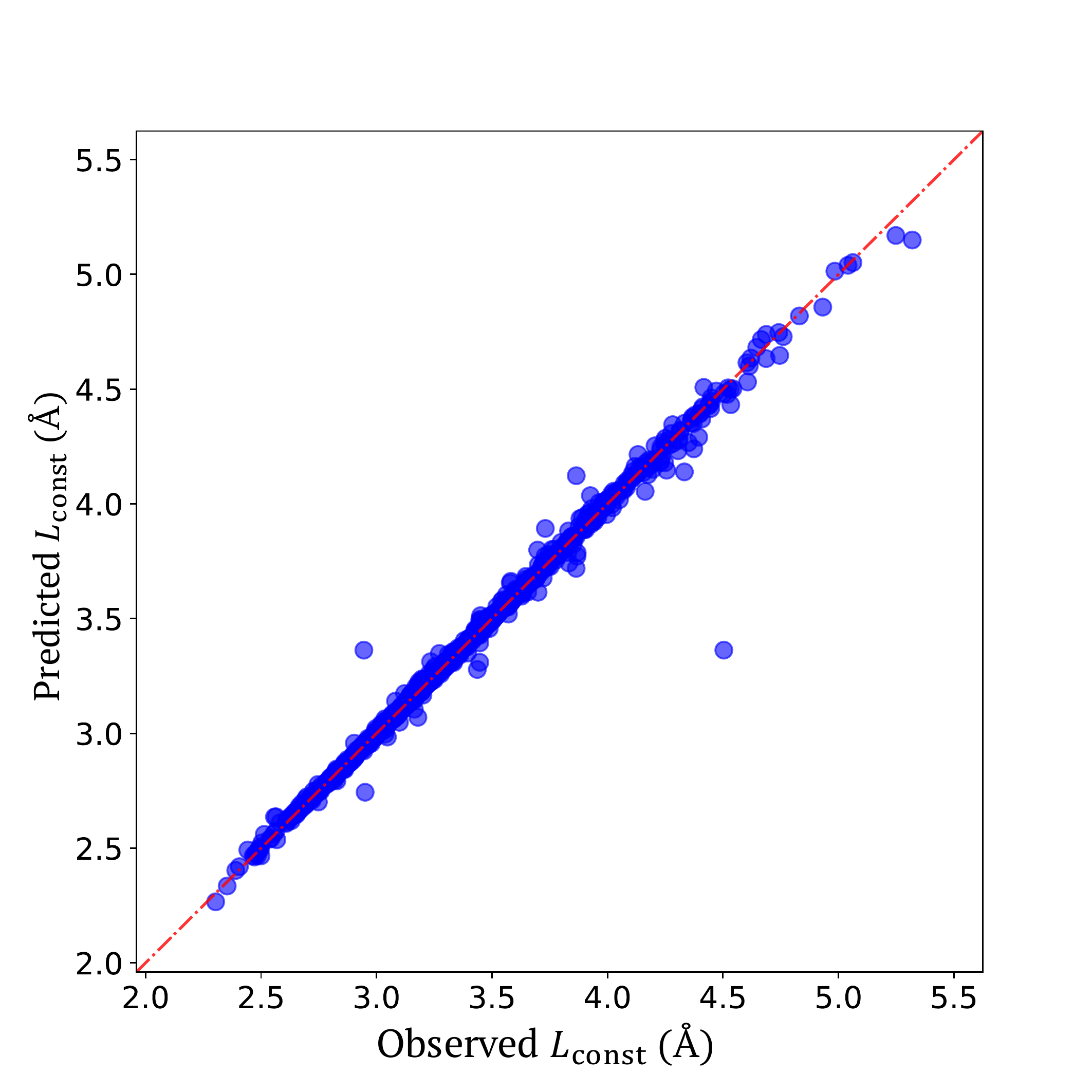} } \label{NlfsBestLatt}}
 \subfloat[][] {{\includegraphics[scale=0.27]{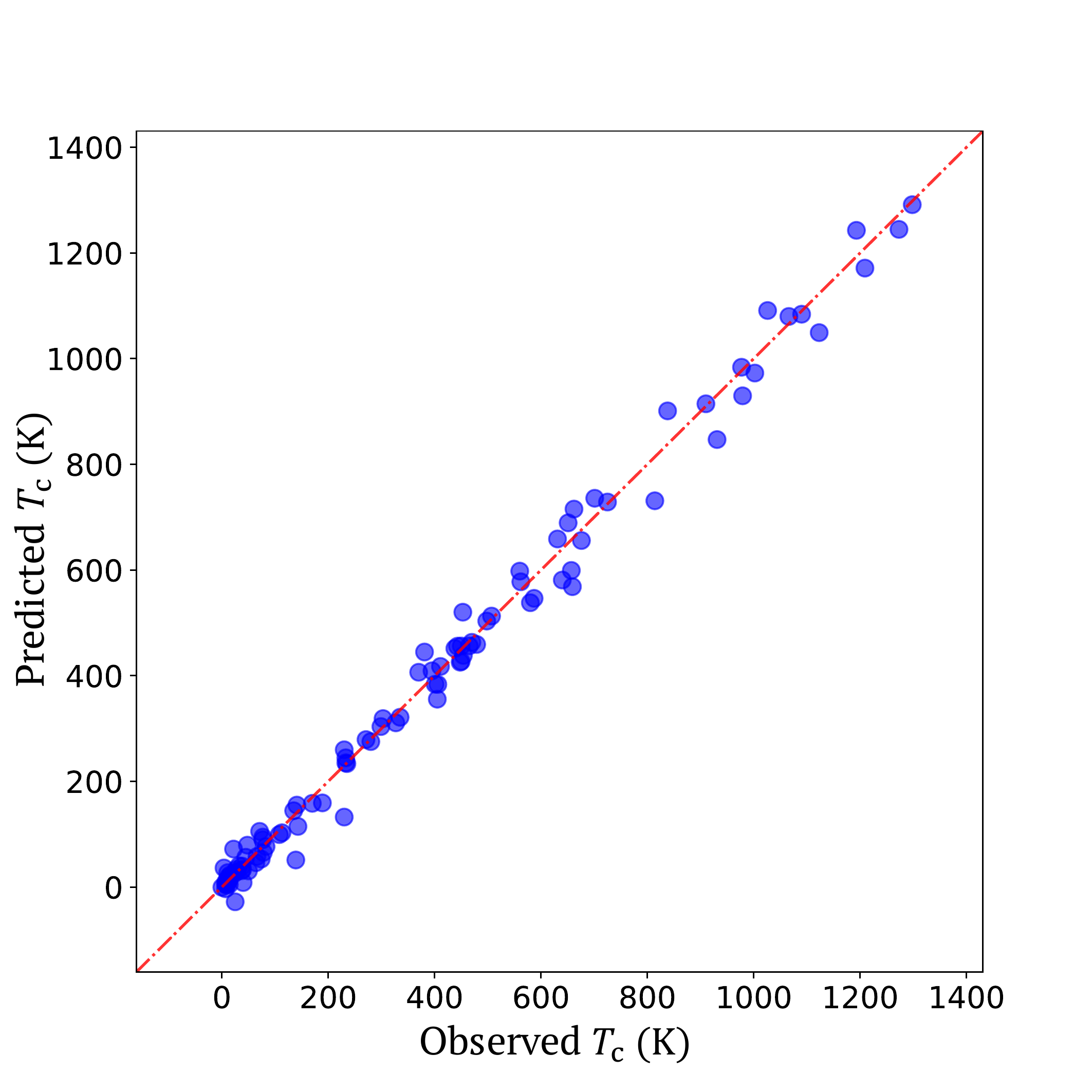} } \label{NlfsBestTc}}
\caption{From left to right, observed and predicted target variable by taking ensemble averaging of 139 ($E_{form}$ problem), 57 ($L_{const}$ problem) and 59 ($T_c$ problem) best prediction models including similarity measure information. Ensemble models yield a $PA$ with $R^2$ scores of 0.982 (MAE: 0.101 eV) for predicting $E_{form}$ problem, 0.992 (MAE: 0.011 \AA) for predicting $L_{const}$ problem and 0.991 (MAE: 24.16 K) for predicting $T_c$ problem.}
\end{figure*}

Figure \ref{ABmap}b shows an enlarged view of the affinity matrix for two groups of typical materials denoted by G1 and G2. We can clearly see that the affinities between materials within each of the two groups, G1 and G2, exceed 0.7, showing high intra-group similarities. In contrast, the affinities between materials in different groups are smaller than 0.2, showing significant dissimilarity between G1 and G2. Further detailed investigation reveals that the materials in G1 are oxide, nitride, and carbide. The maximum common positive oxidation number of the A elements is greater than or equal to the maximum common negative oxidation number of the B elements for the compounds in this group. On the other hand, the materials in G2 are halides of alkaline metal, oxide, nitride, and carbide, for which the maximum common positive oxidation number of the A elements is less than or equal to the maximum common negative oxidation number of the B elements. Further investigation shows that only seven among 24 compounds in G1 have computed electronic structures with a band gap. In contrast, half of the compounds in G2 have computed electronic structures with a band gap. The obtained results suggest that the bonding nature of compounds in G1 is different from that of compounds in G2. 

The linearities between the target variable and the predicting variables for the two groups are summarized in Fig$.$\ref{ABmap}c. The diagonal plots show the correlations between the observed and predicted values for the target variables obtained using linear models of the predicting variables for the materials in the two groups. The off-diagonal plots show the correlations between the observed values and predicted values for the target variables obtained using the linear models of the other groups. We could again confirm the intra-group similarity and the dissimilarity between different groups in terms of the linearity between the target and predicting variables for the compounds in the two groups. 

To quantitatively evaluate the validity of the analysis process, we embedded the similarity measured by the committee machine into the regression of $E_{form}$ of the $\textit{Fm}{\bar{3}}\textit{m}$  AB materials. To predict the value of the target variable for a new material, instead of using the entire available dataset, we used only one-third of the available materials having the highest similarity to the new material. It should again be noted that the similarity between the materials in the dataset and the new material can be determined without knowing the value of the target physical property, using the value predicted by ensemble averaging of the non-linear regression models. 

\begin{table*}
\centering
\caption{$PA$ values for  $E_{form}$, $L_{const}$, and $T_c$ prediction problems. The results obtained with and without using the similarity measure (SM) information  are shown for comparison.}
\begin{tabular}{p{5.0cm}p{1.5cm}p{1.5cm}p{1.5cm}p{1.5cm}p{1.5cm}p{1.5cm}p{1.5cm}}
	\hline\hline
  	\centering
	  \multirow{2}{5.0cm}{Prediction method} &  \multirow{2}{0.7cm}{}& \multicolumn{2}{c}{$E_{form}$ (eV)} & \multicolumn{2}{c}{$L_{const}$ (\AA)} & \multicolumn{2}{c}{$T_{c}$ (K)}  \\ 
 	  \cline{3-8}
  	  & & without SM  & with SM & without SM & with SM & without SM & with SM  \\
  	\hline
    \multirow{2}{5.0cm}{GKR with all variables} & $R^2$ & 0.929 & 0.954 & 0.982  & 0.986 & 0.893 & 0.929 \\
    \cline{2-8}
    & MAE & 0.189 & 0.154 & 0.022 &0.018 & 78.80 &  58.09\\
    \hline
  	\multirow{2}{5.0cm}{GKR with the best variable combination} & $R^2$ & 0.967  & 0.978 & 0.989  & 0.992  & 0.968  & 0.988  \\
  	    \cline{2-8}
  	& MAE & 0.122 & 0.110 & 0.014 & 0.013 & 42.74 & 25.76 \\
  	\hline
  	\multirow{2}{5.0cm}{Ensemble of GKRs with top selected best variable combinations}  & $R^2$ &0.972 & 0.982 & 0.991 &  0.992 & 0.974 & 0.991 \\
  	\cline{2-8}
  	& MAE &0.117 & 0.101 & 0.013 & 0.011 & 37.87 & 24.16\\
	\hline\hline
\end{tabular}
	\label{ConclusionTable}
\end{table*}

Table \ref{ConclusionTable} summarizes the $PA$ in predicting $E_{form}$ values  of the $\textit{Fm}{\bar{3}}\textit{m}$ materials  obtained using several regression models with the designed predicting variables. The non-linear model obtained using ensemble averaging  of the best non-linear regression models, having an $R^{2}$ score of 0.972 (MAE: 0.117 eV), could be improved significantly to an $R^{2}$ score of 0.982 (MAE: 0.101 eV) regarding the information from the similarity measurement (Fig$.$\ref{NlfsBestAB}). Therefore, the obtained results provide significant evidence to support our hypothesis that the similarity voted by the committee machine reflects the similarity in the actuating mechanisms of the target material physical property.

\subsection*{Experiment 2: Mining quantum calculated lattice parameter for body-centered cubic structure data}\label{secLattConst}

In this experiment, a dataset of 1541 binary AB body-centered cubic ($BCC$) crystals with a 1:1 element ratio was collected from Ref.\cite{Takahashi2017}. We focused on the computed lattice constant value $L_{const}$ of the crystals. The A elements corresponded to almost all transition metals \{Ag, Al, As, Au, Co, Cr, Cu, Fe, Ga, Li, Mg, Na, Ni, Os, Pd, Pt, Rh, Ru, Si, Ti, V, W, and Zn\} and the B elements corresponded to those with atomic numbers in the ranges of $1$--$42$, $44$--$57$, and $72$--$83$. This dataset included unrealistic materials such as the binary material AgHe, which incorporates He, an element that is known to possess a closed shell structure and is, therefore, unlikely to form a solid. 

To describe each material, we used a combination of 17 variables that related to basic physical properties of the A and B constituent elements, as summarized in Table \ref{LatTable}. These chosen properties were as follows: the (1) atomic radius ($r_A, r_B$); (2) mass ($m_{A}, m_{B}$); (3) atomic number ($Z_{A}, Z_{B}$); (4) number of electrons in outermost shell ($n_{eA}, n_{eB}$); (5) atomic orbital (${\ell}_A, {\ell}_B$); and (6) electronegativity (${\chi}_{A}, {\chi}_{B}$). The atomic orbital values were converted from categorical symbols $s$, $p$, $d$, $f$ to numerical values representing the orbitals, \ie 0, 1, 2, 3, respectively. To embed the structure information, four more properties were included: (7) the density of atoms per unit volume ($ {\rho}_{A}, {\rho}_{B}$); (8) the unit cell density $\rho$; (9) the difference in electronegativity  $d_{\chi}$; and (10) the sum of the atomic orbital B and difference in electronegativity ${sum}_{AD}$ (see Ref.\cite{Takahashi2017}).  

A kernel regression-based variable selection with 3-times 10-fold cross-validation was performed to examine all combinations of the 17 variables. From the total number of screening variable combinations ($2^{17}$ - 1 = 131,071), we found 60,568 variable combinations for deriving regression models with $R^2$ scores exceeding 0.90 (Fig$.$\ref{NLFS_scores}). Among them, there were 57 variable combinations yielding regression models with $R^2$ scores exceeding 0.9895. The highest $PA$ for this experiment is 0.989 (MAE:  0.014 \AA), which was obtained using the combination $\left\{\rho, {\ell}_{A},  r_{covB}, m_{A}, m_{B}, {\rho}_{B}, n_{eB} \right\}$. We could obtain a better $PA$ with an $R^2$ score of 0.991 (MAE: 0.013 \AA) by taking ensemble averaging of GKR models which derived from the 57 selected variable combinations. This result is a considerable improvement in comparison with the maximum $PA$ ($R^2$ score: 0.90) of the support vector regression technique with the feature selection strategy mentioned in \cite{Takahashi2017}.

\begin{figure*}[t]
\centering
 \includegraphics[scale=0.35]{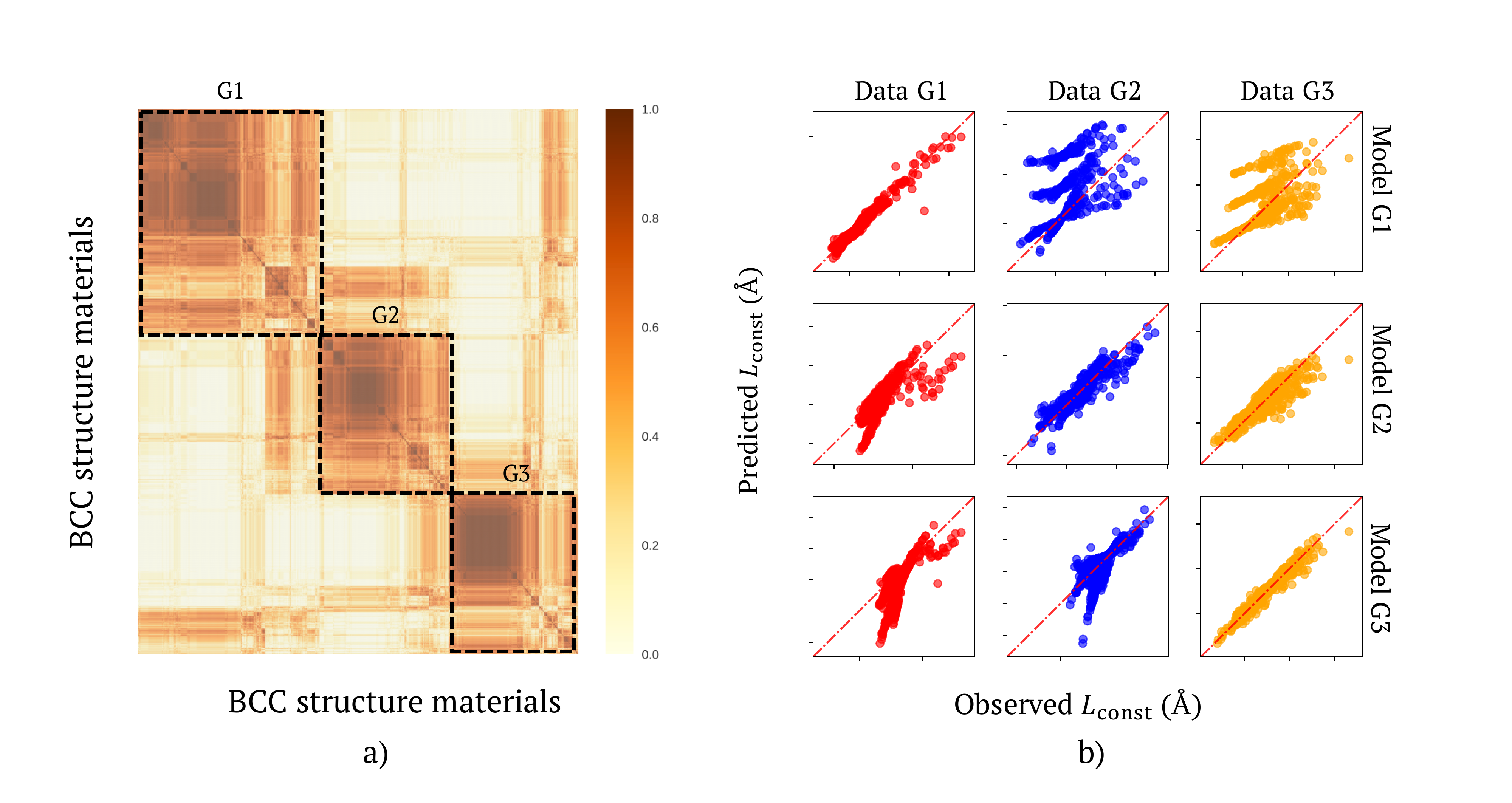}  \label{LattConstConfMt}
  \caption{
a) Similarity matrix between materials for $L_{const}$ prediction problem yielded by regression-based committee voting machine. This similarity matrix can be approximated as three disjoint groups of materials denoted by G1, G2, and G3. b) Confusion matrixes measuring linear similarities among materials in each group, as well as dissimilarities between models generated for materials in different groups.}
  \label{LattConstmap}
\end{figure*}

\begin{table}[t]
\centering
\caption{
Designed predicting variables describing intrinsic properties of constituent elements and structural properties of materials in the lattice parameter prediction problem. A and B are elements of the binary AB $BCC$ materials.
}
\label{LatTable}
\begin{tabular}{p{2.6cm}p{5.5cm}}
	\hline\hline
		Category & Predicting variables \\ 
	\hline
	Atomic properties of metals A & $r_{covA}$, $m_{A}$, $Z_{A}$,  $n_{eA}$, ${\ell}_A$, ${\chi}_{A}$, ${\rho}_{A}$ \\ 
	 \hline
	Atomic properties  of metals B & $r_{covB}$, $m_{B}$, $Z_{B}$,  $n_{eB}$, ${\ell}_B$, ${\chi}_{B}$, ${\rho}_{B}$\\ 
	 \hline
	 Structural \& additional information &$\rho$, $d_{\chi}$, ${sum}_{AD}$ \\
	\hline\hline
\end{tabular}
\end{table}

In the regression-based clustering analysis, the 57 selected variable combinations accompanied by 1000 initial randomized states for each combination were used to search for the most probable clustering results to construct the committee machine. The affinity matrix obtained for all materials is shown in Fig$.$\ref{LattConstmap}a, after rearrangement by a hierarchical clustering algorithm \cite{HAC}. By utilizing this similarity, we could roughly divide all materials in the dataset into three groups: G1, G2, and G3. Further investigation revealed that most materials in G1 are constructed from two heavy transition metals. In contrast, the materials in G2 and G3 are constructed from a metal and a non-metal element, \eg oxide and nitride. For a given A element, the $L_{const}$ of the materials in G1 increases with the atomic number of the B element. On the other hand, the $L_{const}$ of the materials in G2 remains constant for the materials sharing the same A element. Further, the $L_{const}$ for the materials in group G3 mainly depends on the electronegativity difference between the constituent elements A and B. Note that the materials in these three groups are visualized in detail in the Supplemental Materials. The linearities between the observed and predicting variables for these groups are shown in Fig$.$\ref{LattConstmap}b.

To predict the $L_{const}$ of a new material, we applied the same strategy as that explained in the previous experiment. Table \ref{ConclusionTable} summarizes the $PA$ values obtained in our experiments. The non-linear model obtained using ensemble averaging of the best 57 non-linear regression models and having an $R^2$ score of 0.991 (MAE: 0.013 \AA) could be marginally improved to an $R^2$ score of 0.992 (MAE: 0.011 \AA) by including information from the similarity measurement (Fig$.$\ref{NlfsBestLatt}). 

\begin{figure*}[t]
\centering
  \includegraphics[scale=0.3]{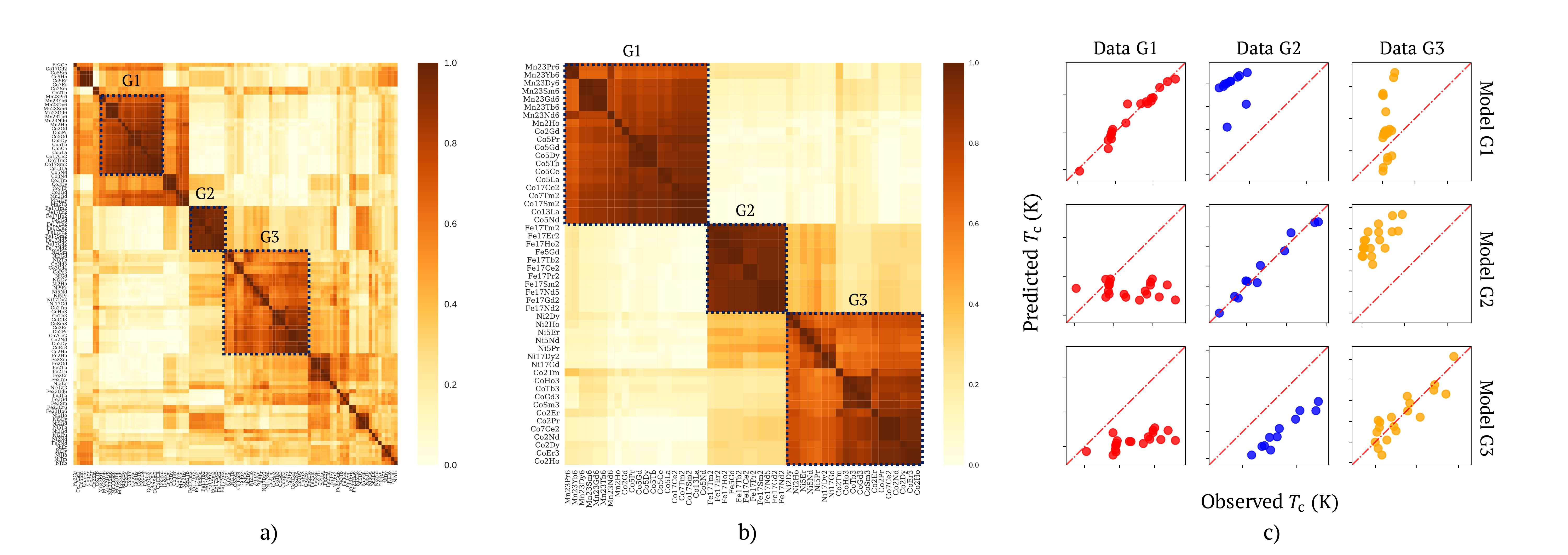}
  \caption{
  a) Similarity matrix between the rare-earth--transition metal alloys yielded by regression-based committee voting machine. b) Enlarged view of highly similar elements in G1, G2, and G3 regions in similarity matrix. c) Confusion matrixes measuring linear similarities among alloys in each group as well as dissimilarities between models generated for alloys in different groups.}
  \label{Tcmap}
\end{figure*}
 
\begin{table}[t]
\centering
\caption{Designed predicting variables describing intrinsic properties of constituent elements and structural properties in $T_{c}$ value prediction for the rare-earth--transition metal alloys problem.}
\label{TcTable}
\begin{tabular}{p{2.6cm}p{5.5cm}}
	\hline\hline
	Category & Predicting variables \\ 
	\hline
	Atomic properties of transition metals & $Z_{T}$, ${r_{cov}}_{A}$, ${IP}_{T}$, ${\chi}_{T}$, $S_{3d}$, $L_{3d}$, $J_{3d}$ \\ 
	 \hline
	Atomic properties of rare-earth metals& $Z_{R}$, ${r_{cov}}_{R}$, ${IP}_{R}$, ${\chi}_{R}$, $ S_{4f}$, $ L_{4f}$, $J_{4f}$, $J_{4f}g_{j}$, $J_{4f}\left(1-g_{j}\right)$ \\ 
	 \hline
	Structural information &  $C_{T}$, $C_{R}$, $r_{TT}$, $r_{TR}$, $r_{RR}$  \\
	\hline\hline
\end{tabular}
\end{table}

\subsection*{Experiment 3: Mining experimentally observed Curie temperature data of rare-earth--transition metal alloys}\label{Tc}
In this experiment, we collected experimental data related to 101 binary alloys consisting of transition and rare-earth metals from the NIMS AtomWork database \cite{paulingfile, atomwork}, which included the crystal structures of the alloys and their observed Curie temperatures $T_c$. 

To represent the structural and physical properties of each binary alloy, we used a combination of 21 variables divided into three categories, as summarized in Table \ref{TcTable}. The first and second categories contained predicting variables describing the atomic properties of the transition metal elements and rare-earth elements, respectively. The properties were as follows: (1) atomic number ($Z_R$, $Z_T$); (2) covalent radius ($r_{covR}$, $r_{covT}$); (3) first ionization (${IP}_R$, ${IP}_T$); and (4) electronegativity (${\chi}_R$, ${\chi}_T$).  In addition, predicting variables related to the magnetic properties were included: the (5) total spin quantum number ($S_{3d}$, $S_{4f}$); (6) total orbital angular momentum quantum number ($L_{3d}$, $L_{4f}$); and (7) total angular momentum ($J_{3d}$, $J_{4f}$). For $R$ metallic elements, additional variables $J_{4f}g_{j}$ and $J_{4f}\left(1-g_{j}\right) $ were added, because of the strong spin-orbit coupling effect. 

As in the two previous experiments, a third category variable was chosen which contained values calculated from the crystal structures of the alloys reported in the AtomWork database \cite{paulingfile, atomwork}. The designed predicting variables included the transition $(C_T)$ and rare-earth $(C_R)$ metal concentrations. Note that, if we use the atomic percentage for the concentration, the two quantities are not independent. Therefore, in this work, we measured the concentrations in units of atoms/\AA$^3$; this unit is more informative than the atomic percentage as it contains information on the constituent atomic size. As a consequence, $(C_T)$ and $(C_R)$ were not completely dependent. Other additional structure variables were also added: the mean radius of the unit cell between two rare-earth  elements $r_{RR}$,  between two  transition metal elements $r_{TT}$, and between transition and rare-earth elements $r_{TR}$. We set the experimentally observed $T_c$ as the target variable.

A kernel regression-based variable selection analysis was performed for these data using leave-one-out cross-validation. Among all the examined variable combinations, ($2^{21}-1 = 2,097,151$), we found 84,870 combinations for which the corresponding GKR models exhibited $R^2$ scores exceeding 0.90 (Fig$.$\ref{NLFS_scores}). Among them, there were 59 variable combinations yielding GKR models associated with $R^2$ scores exceeding 0.95. These predicting variable combinations were selected for the next analysis step. The highest $PA$ in this experiment was 0.968 (MAE: 42.74 K), obtained using the combination $\left\{C_{R}, Z_{R}, Z_{T}, \chi_{T}, r_{covT}, L_{3d}, J_{3d}\right\}$. We could obtain a better $PA$ with an $R^2$ score of 0.974 (MAE: 37.87 K), by applying ensemble averaging to the GKR models, which were derived from the selected 59 variable combinations.  We considered these variable combinations as candidates for the next step of the analysis.

In the regression-based clustering analysis, 59 variable combinations with 1000 initial randomized states were used to search for the most probable clustering results to construct the committee machine to vote for the similarity between the alloys. The obtained affinity matrix for all the alloys is shown in Fig$.$\ref{Tcmap}a. An enlarged view of the three groups of alloys having high similarity (denoted G1, G2, and G3) is shown in Fig$.$\ref{Tcmap}b. Further investigation revealed that G1 includes Mn- and Co-based alloys with high $T_c$, \eg Mn$_{23}$Pr$_6$ (448 K), Mn$_{23}$Sm$_6$ (450 K), Co$_5$Pr (931 K), and Co$_5$Nd (910 K). Other low-$T_c$ Co-based alloys, \eg Co$_2$Pr (45 K) and Co$_2$Nd (108 K), are counted as having higher similarity with Ni-based alloys in G3, \eg Ni$_5$Nd (7 K) and Ni$_2$Ho (16 K). In contrast, G2 includes all the Fe-based $Fe_{17}RE_2$ alloys, where $RE$ represents different rare-earth metals. To confirm the value of our similarity measure, Fig$.$\ref{Tcmap}c shows the linearities between the observed and predicting variables for these groups, as well as the dissimilarities among these groups. 

In the next analysis step, we utilized the obtained similarity measure to predict $T_c$ for a new material by using the same strategy used in the two previous experiments.  The non-linear model obtained using ensemble averaging of the best non-linear regression models and having an $R^2$ score of 0.974 (MAE: 37.87 K) could be improved significantly to attain an $R^2$ score of 0.991 (MAE: 24.16 K) by utilizing the information from the similarity measurement (Fig$.$\ref{NlfsBestTc} and Table \ref{ConclusionTable}). The obtained results provide significant evidence to support our hypothesis that the similarity voted for by the committee machine indicates the similarity in the actuating mechanisms of the $T_c$ of the binary alloys.

\section{Conclusion}
In this work, we proposed a method to measure the similarities between materials, focusing on specific physical properties, to describe and interpret the actual mechanism underlying a physical phenomenon in a given problem. The proposed method consists of three steps: variable evaluation based on non-linear regression, regression-based clustering, and similarity measurement with a committee machine constructed from the clustering result. Three datasets of well-characterized crystalline materials represented by key atomic predicting variables were used as test beds. The formation energy, lattice parameter, and Curie temperature were considered as target physical properties of the examined materials. Our experiments show that rational and meaningful group structures can be obtained with the help of the proposed approach. The similarity measure information helped significantly increase the prediction accuracy for the material physical properties. Through use of ensemble top kernel ridge prediction models, the $R^2$ score increased from 0.972 to 0.982 for the formation energy prediction problem; 0.991 to 0.992 for the lattice constant prediction problem, and 0.974 to 0.991 for the Curie temperature prediction problem after utilizing the similarity information. Thus, our results indicate that our  proposed data analysis flow can systematically facilitate further understanding of a given phenomenon by identifying similarities among materials in the problem dataset. 

\begin{acknowledgments}
This work was partly supported by PRESTO and by the “Materials Research by Information Integration Initiative” (MI$^2$I) project of the Support Program for Start-Up Innovation Hub, from the Japan Science and Technology Agency (JST), and by JSPS KAKENHI Grant-in-Aid for Young Scientists (B) (Grant JP17K14803), Japan. 
\end{acknowledgments}

\section*{Author contributions statement}
Duong -Nguyen Nguyen and Hieu-Chi Dam conceived the experiments;  Duong-Nguyen Nguyen, Viet-Cuong Nguyen, and Tuan-Dung Ho conducted the experiments; and Duong-Nguyen Nguyen, Tien-Lam Pham, Truyen Tran, Keisuke Takahashi, and Hieu-Chi Dam analyzed the results. Duong-Nguyen Nguyen and Hieu-Chi Dam wrote the paper, and all authors reviewed the manuscript. 

T$  $he authors declare no competing interests.

\bibliography{main_IUCrJ}

\end{document}